\documentclass[letterpaper,preprintnumbers,prd,twocolumn,nofootinbib,nobibnotes,showpacs]{revtex4}
\usepackage{amsfonts}
\usepackage{mathrsfs}
\usepackage{epsfig}
\usepackage{graphicx}%
\usepackage{dcolumn}
\usepackage{amsmath}

\makeatletter
\def\btt#1{\texttt{\@backslashchar#1}}%
\DeclareRobustCommand\bblash{\btt{\@backslashchar}}%
\makeatother
\begin{document}

\title{When scalar field is kinetically coupled to the Einstein tensor}
\author{Changjun Gao}\email{gaocj@bao.ac.cn}\affiliation{The
National Astronomical Observatories, Chinese Academy of Sciences}
\affiliation{{Key Laboratory of Optical Astronomy, NAOC, CAS,
Beijing, 100012
}}
\affiliation{{Kavli Institute for Theoretical
Physics China, CAS, Beijing 100190, China }}
\date{\today}

\date{\today}

\begin{abstract}
We explore the cosmic evolution of a scalar field with the kinetic term coupled to the Einstein tensor.
We find that, in the absence of other matter sources or in the presence of only pressureless matter, the scalar behaves as pressureless matter and the sound speed of the scalar is vanishing. These properties enable the scalar field to be a candidate of cold dark matter. By also considering the scalar potential, we find the scalar field may play the role of both dark matter and dark energy. In this case, the equation of state of the scalar can cross the phantom divide, but this can lead to the sound speed becoming superluminal as it crosses the divide, and so is physically forbidden. Finally, if the kinetic term is coupled to more than one Einstein tensor, we find the equation of state is always approximately equal to -1 whether the potential is flat or not, and so the scalar may also be a candidate for the inflaton.
\end{abstract}


\maketitle

\section{Introduction}
Although at present there are no known fundamental scalar particles in nature, scalar fields play an important role in physics and cosmology. In physics, scalar fields are present in Jordan-Brans-Dicke theory  as Jordan-Brans-Dicke scalar \cite{brans:1961}; in Kaluza-Klein compactification theory as the radion \cite{csaki:2000}, in the Standard Model of particle physics as the Higgs boson \cite{higgs:1964}, in the low-energy limit of the superstring theory as the dilaton \cite{gibbons:1985} or tachyon \cite{sen:2002} and so on. In cosmology, scalar fields are present as the inflaton \cite{guth:1981} to drive the inflation of the early Universe and currently as the quintessence \cite{{ratra:1988},{caldwell:1998},zlatev:1999} or phantom \cite{caldwell:1999} fields, driving the acceleration the Universe.

In general, the action of scalar-tensor theories of gravity are given by
\begin{eqnarray}
S&=&\int d^4x\sqrt{-g}\left[f\left(\phi,R,\ R_{\mu\nu}R^{\mu\nu},\ R_{\mu\nu\lambda\gamma}R^{\mu\nu\lambda\gamma}\right)
\right.\nonumber\\&&\left.+K\left(\phi,\ \partial_\mu\phi\partial^{\mu}\phi,\ \nabla^2\phi\right)+V\left(\phi\right)+S_m\right]\;,
\end{eqnarray}
where $\phi$ is the scalar field, $V(\phi)$ the scalar potential and $R$ the Ricci scalar. Here $f$ and $K$ are arbitrary functions of
the corresponding variables. These theories cover the $f(R)$ modified gravity \cite{capo:2003}, the Gauss-Bonnet gravity \cite{cvetic:2002}, the quintessence scalar, the phantom scalar, the quintom fields \cite{feng:2004},
the K-essence scalar theory \cite{picon:2001}, the tachyon, dilaton, and so on. We note that these theories have already been investigated extensively.

On the contrary, to our knowledge, the even more general scalar tensor theories of gravity such as
\begin{eqnarray}
&&S=\int d^4x\sqrt{-g}\left[f\left(\phi,R,\ R_{\mu\nu}R^{\mu\nu},\
R_{\mu\nu\lambda\gamma}R^{\mu\nu\lambda\gamma}\right)
\right.\nonumber\\&&\left.+K\left(\phi,\
\partial_\mu\phi\partial^{\mu}\phi,\ \nabla^2\phi,\
R^{\mu\nu}\partial_{\mu}\phi\partial_{\nu}\phi,\ \phi
R^{\mu\nu}\nabla_{\mu}\partial_{\nu}\phi,\cdot\cdot\cdot\right)\right.\nonumber
\\
&&\left.+V\left(\phi\right)+S_m\right]\;,
\end{eqnarray}
are has not been so well investigated. The new coupling between the derivative of scalar and the spacetime curvature may appear in some Kaluza-Klein
theories \cite{shafi:1985,shafi:1987,linde:1990}. In 1993, Amendola \cite{amendola:1993} studied the scalar-tensor theory with the Lagrangian linear in the Ricci scalar $R$, quadratic in $\phi$, and containing terms as follows:
\begin{eqnarray}
&&R\partial_{\mu}\phi\partial^{\mu}\phi\;,\ \ \
R_{\mu\nu}\partial^{\mu}\phi\partial^{\nu}\phi\;,\ \ \
R\phi\nabla^2\phi\;,\ \ \nonumber\\&& R_
{\mu\nu}\phi\nabla^{\mu}\partial^{\nu}\phi\;,\ \ \
\partial_{\mu}R\partial^{\mu}\phi\;,\ \ \ \nabla^2R\phi\;.\ \ \
\end{eqnarray}
Amendola \cite{amendola:1993} investigated a cosmological model with the
only derivative coupling term $R_{\mu\nu}\partial^{\mu}\phi\partial^{\nu}\phi$ and presented some analytical inflationary
solutions. A general model containing $R\partial_{\mu}\phi\partial^{\mu}\phi$ and $R_{\mu\nu}\partial^{\mu}\phi\partial^{\nu}\phi$ has been discussed by Capozziello et al., \cite{Capozziello:1999,Capozziello:2000}. They showed
that the de Sitter spacetime is an attractor solution in the model. In 2007, Daniel and
Caldwell \cite{daniel:2007} studied a theory with the derivative coupling term of $R_{\mu\nu}\partial^{\mu}\phi\partial^{\nu}\phi$ and
found the constraints on the coupling parameter with Solar system tests.

In general, these scalar-tensor theories give both the Einstein equations and the equation of motion for the scalar in the form of fourth-order differential equations. However, recently Sushkov \cite{sushkov:2009} showed that in the case of the kinetic term only coupled to the Einstein tensor, the equation of motion for the scalar is reduced to second order. Thus, from the point of view of physics, this theory can be interpreted as a ``good'' theory. The reason for this is very simple. It is well understood that  tensors constructed from the metric tensor and its derivatives (up to second order), only the metric tenor $g_{\mu\nu}$ and the Einstein tensor $G_{\mu\nu}$ are divergence free:
\begin{eqnarray}
g_{\mu\nu}^{\ \ ;\nu}=0\;, \ \ G_{\mu\nu}^{\ \ ;\nu}=0\;.
\end{eqnarray}
Therefore for the Lagrangian
\begin{eqnarray}
\mathscr{L}=\left(\frac{\varepsilon}{2}g^{\mu\nu}+\frac{\alpha}{2} G^{\mu\nu}\right)\partial_{\mu}\phi\partial_{\nu}\phi+V\left(\phi\right)\;,
\end{eqnarray}
with $\varepsilon$ and $\alpha$ constants,  the equation of motion takes the form of
\begin{eqnarray}
\left(\varepsilon g^{\mu\nu}+\alpha
G^{\mu\nu}\right)\nabla_{\mu}\partial_{\nu}\phi-V^{'}=0\;,
\end{eqnarray}
 which is a second order differential equation.

In this paper, we will investigate the cosmic evolution of a scalar field with the kinetic term coupled to more than one Einstein tensor. We find this scalar field presents us with two very interesting characteristics. When the kinetic term is coupled to only one Einstein tensor, and in the absence of any other matter sources or in the presence of only pressureless matter, the scalar behaves exactly as the pressureless matter. Thus it could be a candidate for cold dark matter. On the other hand, when the kinetic term is coupled to more than one Einstein tensor, the scalar field can have the equation of state $w\simeq-1$ over the whole history of the Universe. Therefore, the scalar  field can play the role of a dynamic cosmological constant.

The paper is organized as follows. In section II, we shall investigate the cosmic evolution of the scalar with the kinetic term coupled to one Einstein tensor. We find it to be a possible candidate for either cold dark matter or both cold dark matter and dark energy. In section III, we investigate the cosmic evolution of the scalar with the kinetic term coupled to more than one Einstein tensor. Section IV gives the conclusions and discussion. We shall use the system of units with $G=c=\hbar=k=1$ and the metric signature $(-,\ +,\ +,\ +)$ throughout the paper.
\section{Coupled to One Einstein tensor}
\subsection{Equations of Motion}
The Lagrangian density of quintessence-phantom (or quintom) is given by
\begin{eqnarray}
\mathscr{L}_{\textrm{QP}}=\frac{\varepsilon}{2}g^{\mu\nu}\partial_{\mu}\phi\partial_{\nu}\phi+V\left(\phi\right)\;,
\end{eqnarray}
where $g^{\mu\nu}$ is the metric tensor. Here $\varepsilon=+1$ corresponds to quintessence and $\varepsilon=-1$ to a phantom field.
Let's investigate the scalar field with the lagrangian density as follows
 \begin{eqnarray}
\mathscr{L}=\frac{\alpha}{2}G^{\mu\nu}\partial_{\mu}\phi\partial_{\nu}\phi+V\left(\phi\right)\;,
\end{eqnarray}
where
\begin{eqnarray}
G^{\mu\nu}=R^{\mu\nu}-\frac{1}{2}g^{\mu\nu}R\;,
\end{eqnarray}
is the Einstein tensor. $\alpha$ is assumed to be a positive constant. We note that differently from quintessence or phantom fields,
the kinetic term of this scalar field is coupled not to the metric tensor, but to the Einstein tensor. This modification leads to the scalar field
behaving as the Einstein cosmological constant in Minkowski spacetime due to the fact that $G^{\mu\nu}=0$.
Then from the action
\begin{eqnarray}
\label{eq:999}
S=\int d^4x\sqrt{-g}\left(\frac{R}{16\pi}+\mathscr{L}\right)+S_m\;,
\end{eqnarray}
we obtain the Einstein equations
\begin{eqnarray}
G_{\mu\nu}&=&8\pi T_{\mu\nu}+8\pi Vg_{\mu\nu}+8\pi \alpha\big\{-{\textstyle\frac12}\nabla_\mu\phi\,\nabla_\nu\phi\,R
\nonumber\\
&& +2\nabla_\beta\phi\,\nabla_{(\mu}\phi R^\beta_{\nu)}
+\nabla^\sigma\phi\,\nabla^\beta\phi\,R_{\mu\sigma\nu\beta}\nonumber\\
&&+\nabla_\mu\nabla^\sigma\phi\,\nabla_\nu\nabla_\sigma\phi
-\nabla_\mu\nabla_\nu\phi\,\square\phi\nonumber\\
&&-{\textstyle\frac12}(\nabla\phi)^2
G_{\mu\nu}
+g_{\mu\nu}\big[-{\textstyle\frac12}\nabla^\sigma\nabla^\beta\phi\,\nabla_\sigma\nabla_\beta\phi
\nonumber\\
&&+{\textstyle\frac12}(\square\phi)^2
-\nabla_\sigma\phi\,\nabla_\beta\phi\,R^{\sigma\beta}
\big]\big\}\;,
\end{eqnarray}
and the equation of motion for the scalar
\begin{eqnarray}
\alpha G^{\mu\nu}\nabla_{\mu}\partial_{\nu}\phi-V^{'}=0\;.
\end{eqnarray}
Here $T_{\mu\nu}$ is the energy-momentum tensor for the matter fields derived from the action $S m$. Prime denotes the derivative with respect to
$\phi$. Due to the Bianchi identities,
\begin{eqnarray}
G^{\mu\nu}_{\ \ ;\nu}=0\;,
\end{eqnarray}
the equation of motion for the scalar field is equivalent to the conservation equation for other matter fields
\begin{eqnarray}
T^{\mu\nu} {\ \ ;\nu}=0\;.
\end{eqnarray}

 We note that the equations of motion for a quintessence (or phantom) field is given by
\begin{eqnarray}
\varepsilon g^{\mu\nu}\nabla_{\mu}\partial_{\nu}\phi-V^{'}=0\;.
\end{eqnarray}
The gravitational field described the Solar system is the Schwarzschild solution which obeys
\begin{eqnarray}
G^{\mu\nu}=0\;.
\end{eqnarray}
We conclude that as a test scalar, our Einstein tensor-coupled scalar field would behave as the Einstein cosmological constant $V=const$ in the Solar system, which is not the case for
quintessence-phantom field.

\subsection{As Dark Matter}
In this subsection, we investigate whether the scalar can be a CDM candidate. Consider a spatially flat Friedmann-Robertson-Walker (FRW) Universe
\begin{eqnarray}
 ds^2 = -dt^2+a^2\left(dr^2+r^2d\Omega^2\right)\;,
\end{eqnarray}
where $a$ is the scale factor. We model all other matter sources present in the Universe as perfect fluids.
These matter sources can be baryonic matter, relativistic matter and dark energy. We assume there is no interaction between the scalar field
and other matter fields, other than by gravity. Then the Einstein equations are given by

\begin{eqnarray}
3H^2&=&8\pi\left(\frac{9}{2}\alpha H^2\dot{\phi}^2+\sum\rho_i\right)\;,\nonumber\\
2\dot{H}+3H^2&=&-8\pi\left[-\frac{\alpha}{2}\left(2\dot{H}+3H^2\right)\dot{\phi}^2\right.\nonumber
\\&&\left.-2\alpha H\dot{\phi}\ddot{\phi}+\sum p_{i}\right]\;.
\end{eqnarray}
We define
\begin{eqnarray}
\rho=\sum\rho_i\;,\ \ \  p=\sum p_i\;,
\end{eqnarray}
where $\rho_i,\ \ p_i$ are the energy density and pressure of the i-th matter.

The equation of motion for the scalar can be derived from above Einstein equations Eqs.~(18) or from its equation of motion Eq.~(12) as follows
 \begin{eqnarray}
H\ddot{\phi}+\left(2\dot{H}+3H^2\right)\dot{\phi}=0\;,
\end{eqnarray}
from which we obtain
\begin{eqnarray}
\dot{\phi}=\frac{c_0}{H^2a^3}\;,
\end{eqnarray}
where $c_0$ is an integration constant. So the energy density of the scalar is
\begin{eqnarray}
\rho_{\phi}=\frac{9\alpha}{2}\cdot\frac{c_0^2}{H^2a^6}\;.
\end{eqnarray}
The equation reveals that: in the stiff matter dominated Universe, namely $H^2\propto 1/a^6$, the density
of the scalar is nearly a constant; in the relativistic matter dominated Universe, namely $H^2\propto 1/a^4$, the energy density of the scalar scales as $\rho_{\phi}\propto 1/a^2$ and the scalar has the equation of state $-1/3$ just like the curvature term; in the baryon matter dominated Universe, namely $H^2\propto1/a^3$, the scalar field behaves as pressureless matter; in the cosmological constant dominated Universe, namely $H^2=const$, the energy density of the scalar scales as $\rho_{\phi}\propto 1/a^6$ and so has the equation of state $+1$.

Now we have a Friedmann equation as follows
\begin{eqnarray}
3H^2&=&8\pi\left(\frac{9\alpha c_0^2}{2H^2a^6}+\rho\right)\;.
\end{eqnarray}
By setting
\begin{eqnarray}
\xi^2&\equiv&\frac{27\alpha c_0^2}{16\pi}\;,
\end{eqnarray}
the Friedmann equation can be rewritten as
\begin{eqnarray}
3H^2&=&8\pi\left(\rho+\frac{1}{2}\sqrt{\rho^2+\frac{4\xi^2}{a^6}}-\frac{1}{2}\rho\right)\;.
\end{eqnarray}
The energy density of the scalar turns out to be

\begin{eqnarray}
\rho_{\phi}&=&\frac{1}{2}\left(\sqrt{\rho^2+\frac{4\xi^2}{a^6}}-\rho\right)\;.
\end{eqnarray}
We conclude that, in the absence of all other matter ($\rho=0$)
or in the presence of baryonic matter ($\rho\propto 1/a^3$), the scalar behaves exactly as pressureless dark matter with $\rho {\phi}\propto 1/a^3$. This enables us to propose this scalar field as a candidate for dark matter. Assuming that the energy density of the other matter fields is given by
\begin{eqnarray}
\rho&=&\frac{\rho_{r0}}{a^4}+\frac{\rho_{b0}}{a^3}+\lambda\;,
\end{eqnarray}
where $\rho_{r0},\ \ \rho_{b0}$ are the present-day energy density of relativistic matter and baryon, respectively, and $\lambda$ is the Einstein cosmological constant. Then the pressure and the equation of state for the scalar take the form of
\begin{eqnarray}
p_{\phi}&=&\frac{\rho_{\phi}}{\sqrt{\rho^2+\frac{4\xi^2}{a^6}}}\left(\lambda-\frac{1}{3}\cdot\frac{\rho_{r0}}{a^4}\right)\;.
\end{eqnarray}

\begin{eqnarray}
w&=&\frac{1}{\sqrt{\rho^2+\frac{4\xi^2}{a^6}}}\left(\lambda-\frac{1}{3}\cdot\frac{\rho_{r0}}{a^4}\right)\;.
\end{eqnarray}
We see again the scalar behaves exactly as pressureless matter in the presence of only baryon matter.
Let $\rho 0$ denote the present-day energy density of the Universe. We define
\begin{eqnarray}
&&\Omega_{r0}\equiv\frac{\rho_{r0}}{\rho_0}\;,\ \ \ \Omega_{b0}\equiv\frac{\rho_{b0}}{\rho_0}\;,\nonumber\\&& \Omega_{\lambda}\equiv\frac{\lambda}{\rho _0}\;,\ \ \ k\equiv\frac{\xi}{\rho_0}\;,\ \ \ \Omega\equiv\frac{\rho}{\rho_0}\;.
\end{eqnarray}
Then the equation of state can be written in the dimensionless form
\begin{eqnarray}
w&=&\frac{1}{\sqrt{\Omega^2+\frac{4k^2}{a^6}}}\left(\Omega_{\lambda}-\frac{1}{3}\cdot\frac{\Omega_{r0}}{a^4}\right)\;.
\end{eqnarray}
Observations constrain $\Omega_{b0}=0.04\;,\ \ \
\Omega_{r0}=8.1\cdot10^{-5}\;,\ \ \ \Omega_{\lambda}=0.75\;$. We
find $k=0.458$ corresponding to the density fraction of scalar
${\rho_{\phi0}}/{\rho 0}=0.21$ in the present-day Universe
\cite{komatsu:2009}.

\begin{figure}
\includegraphics[width=6.5cm]{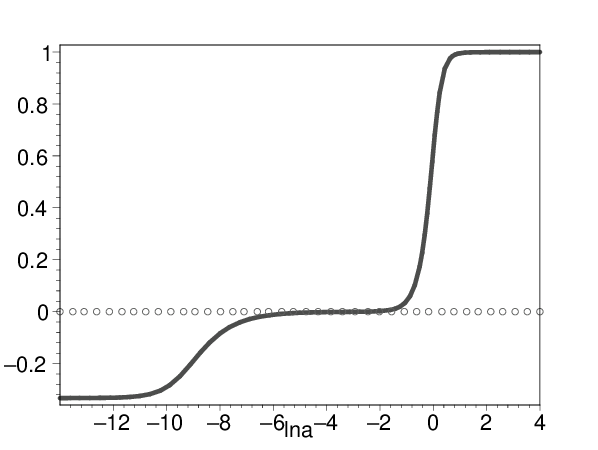}
\caption{\label{fig:eos}The equation of state for the scalar (solid line) and the cold dark matter (circled line). The scalar
behaves as a curvature term ($w=-1/3$), pressureless matter ($w=0$) and stiff matter ($w=1$)
 in radiation, scalar matter and cosmological constant dominated epoch, respectively.  }
\end{figure}

\begin{figure}
\includegraphics[width=6.5cm]{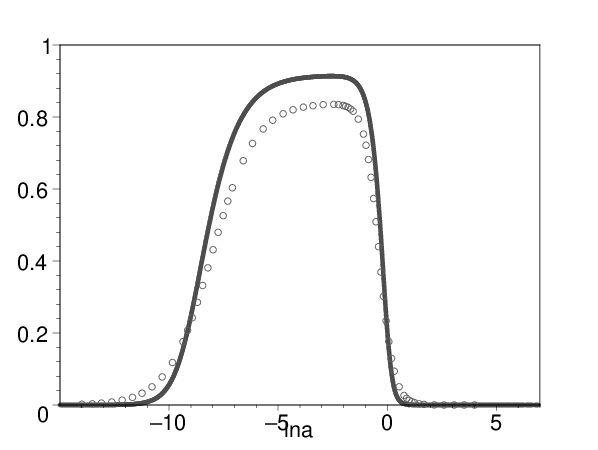}
\caption{ \label{fig:densityfrac}The evolution of density fractions $\frac{\rho_{\phi}}{\rho_{\phi}+\rho}$ of the scalar field (solid line) and $\rho_ d/\rho$ of the cold dark matter (circled line) in the $\Lambda\textrm{CDM}$ model. The scalar contributes significantly only at the interval of $\ln a=-9$ to $\ln a=0$. }
\end{figure}

\begin{figure}
\includegraphics[width=6.5cm]{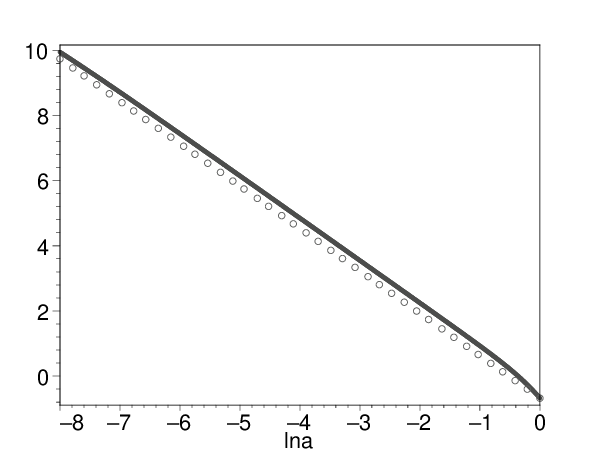}
\caption{\label{fig:densdiff}The evolution of energy density $\log{\rho_{\phi}}$ of the scalar field (solid line) and $\log\rho_d$ of the cold dark matter (circled line) in the $\Lambda\textrm{CDM}$ model. The difference is observationally allowed.}
\end{figure}

In Fig.~\ref{fig:eos}, we plot the equation of state for the scalar. It shows that the scalar behaves as a curvature term ($w=-1/3$), pressureless matter ($w=0$) and stiff matter ($w=1$) in the relativistic matter, pressureless matter and cosmological constant dominated epoch, respectively. In Fig.~\ref{fig:densityfrac}, we plot the density fraction of the scalar during the evolution of the Universe. It is found that the scalar makes significant contribution only at the interval of $\ln a=-9$ to $\ln a=0$. This is not surprising since the scalar only tracks the pressureless matter. The circled line is for $\Lambda\textrm{CDM}$ model. The difference between the two models is observationally allowed as shown in Fig.~\ref{fig:densdiff}.

\subsection{As Dark Matter and Dark Energy}
In the previous subsection, we have shown the scalar can play the role of dark matter. In this subsection, we study whether the scalar can play the role of both dark matter and dark energy by taking account the scalar potential. Since $\alpha$ is assumed to be a positive constant, we may rescale $\phi$ such that the Lagrangian density can be rewritten as
\begin{eqnarray}
\mathscr{L}=\frac{1}{18}G^{\mu\nu}\partial _{\mu}\phi\partial_{\nu}\phi+V\left(\phi\right)\;.
\end{eqnarray}
Then the Einstein equations and the equation of motion take the form of

\begin{eqnarray}
&&3H^2=8\pi\left(\frac{1}{2}H^2\dot{\phi}^2+V+\rho_b+\rho_r\right)\;,\nonumber\\
&&2\dot{H}+3H^2=-8\pi\left[\frac{1}{6}\left(2\dot{H}+3H^2\right)\dot{\phi}^2+\frac{2}{3}\cdot\frac{V^{'}}{H}\dot{\phi}\right.\nonumber
\\&&\left.-V+p_r \right] \;,\nonumber\\
&&H^2\ddot{\phi}+H\left(2\dot{H}+3H^2\right)\dot{\phi}+3V^{'}=0\;,
\end{eqnarray}
where $\rho_b,\ \rho_r$ are the energy density of baryon matter and relativistic matter. $p_r$ is the pressure of relativistic matter.
Now we investigate the evolution of the scalar field. We introduce the following dimensionless quantities
\begin{eqnarray}
&& x\equiv\sqrt{\frac{4\pi}{3}}\dot{\phi}\;,\ \ \ y\equiv\sqrt{\frac{8\pi V}{3H^2}}\;,\ \ \ z\equiv\sqrt{\frac{8\pi \rho_b}{3H^2}}\;,\ \ \ \nonumber\\&&u\equiv-\frac{V^{'}}{\sqrt{8\pi}VH}\;,\ \ \ \Gamma\equiv\frac{V^{''}V}{V^{'2}}\;,\ \ \   N\equiv\ln a\;.
\end{eqnarray}
Here $x^2$ and $y^2$ represent the density parameters of the kinetic and potential terms respectively. Then the above equations can be written in the following autonomous form
\begin{eqnarray}
\label{autoquin1} \frac{dx}{d N}&=&
\frac{\sqrt{6}}{2}y^2u+\frac{x}{1+x^2}\left[-2\sqrt{6}xy^2u \right.\nonumber
\\
&&\left.-3y^2+\left(1-x^2-y^2-z^2\right)\right]\;,\nonumber\\
\frac{dy}{d N} &=&
-\frac{\sqrt{6}}{2}xyu+\frac{3}{2}y+\frac{y}{2\left(1+x^2\right)}\left[-2\sqrt{6}xy^2u\right.\nonumber
\\
&&\left.-3y^2+\left(1-x^2-y^2-z^2\right)\right]\,,\nonumber\\
\frac{dz}{dN}&=&\frac{z}{2\left(1+x^2\right)}\left[-2\sqrt{6}xy^2u-3y^2\right.\nonumber
\\
&&\left.+\left(1-x^2-y^2-z^2\right)\right]\;,\nonumber\\
\frac{du}{dN}&=&-\sqrt{6}\Gamma xu^2+\sqrt{6}xu^2+\frac{u}{2\left(1+x^2\right)}\left[-2\sqrt{6}xy^2u\right.\nonumber
\\
&&\left.-3y^2+\left(1-x^2-y^2-z^2\right)\right]+\frac{3u}{2}\;,
\end{eqnarray}
together with a constraint equation
\begin{equation}
x^2+y^2+z^2+\frac{8\pi \rho_r}{3H^2}=1\,.
\end{equation}
The equation of state $w$ and the
fraction of the energy density $\Omega {\phi}$ for the scalar field
are
\begin{eqnarray}
w&\equiv&\frac{p_\phi}{\rho_\phi}=\frac{{x^2}}{\left(1+x^2\right)\left(3x^2+3y^2\right)}\left[2\sqrt{6}xy^2u
+3y^2\right.\nonumber
\\
&&\left.-\left(1-x^2-y^2-z^2\right)\right]-\frac{2\sqrt{6}xy^2u+3y^2}{3x^2+3y^2}\;,\nonumber\\
\Omega_{\phi}&\equiv&\frac{8\pi\rho_\phi}{3H^2}=x^2+y^2\;.
\end{eqnarray}
As an example, we have considered the exponential potential $V=e^{-\zeta\phi}$ with $\zeta$ a positive constant. Physically, the scalar
 would roll down the potential. Therefore we have $x\geq0,\ y\geq0,\ z\geq0,\ u\geq0$.
We put $\Omega_{b0}=0.04\;,\ \ \ \Omega_{r0}=8.1\cdot10^{-5}\;,\ \ \ \Omega_{\lambda}=0.75\;$.

In Table \ref{table:crit0}, we present the properties of the critical points for the exponential potential. The point (a) corresponds to the relativistic matter dominated epoch and the point is unstable. In this epoch, the scalar has the equation of state $-1/3$. The line (b) corresponds to the scalar(kinetic energy)-plus-baryon dominated epoch and it is a saddle line. In this epoch, the scalar field behaves as dust with equation of state $w=0$. Point (c) corresponds to the potential energy dominated epoch. It is stable and thus an attractor. In the epoch of (c), the expansion of the universe accelerates.

\begin{figure}
\includegraphics[width=6.5cm]{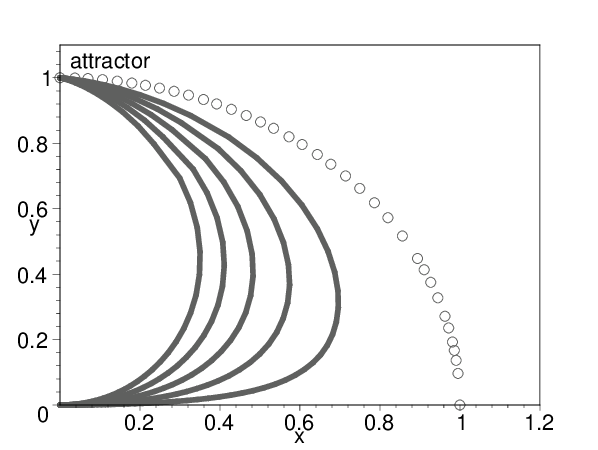}
\caption{\label{fig:exponpot} The phase plane for the exponential potential. The point (0, 0) corresponds to the radiation dominated epoch and the circled arc ($x^2+y^2=1$) corresponds to the scalar dominated epoch. The point (0, 0) is unstable and the point (0 ,1) is stable and an attractor. These trajectories show that the Universe always evolves from the radiation dominated epoch to the scalar
field dominated epoch and ends at the scalar potential dominated epoch.}
\end{figure}

In Fig.~\ref{fig:exponpot}, we plot the phase plane for the scalar with a range of different initial conditions. The point (0, 0) corresponds to the radiation dominated epoch and the circled arc ($x^2+y^2=1$) corresponds to the scalar dominated epoch. The point (0, 0) is unstable and the point (0 ,1) is stable and an attractor. These trajectories show that the Universe always evolves from the radiation dominated epoch to the scalar field dominated epoch and ends at the scalar potential dominated epoch.

\begin{table*}[t]
\begin{center}
\begin{tabular}{|c|c|c|c|c|c|c|c|c|}
Name &  $x$ & $y$ & $z$ & $u$ & Existence & Stability & $\Omega_\phi$
 & $w$ \\
\hline \hline (a) & 0 & 0 & 0 & 0 & All $\zeta$& Unstable node
&   0 & -1/3 \\
\hline (b) & $\sqrt{1-z^2}$ & 0 & z & 0 & All $\zeta$ & Saddle line & ${1-z^2}$ & 0  \\
\hline (c) & 0&1
& 0& $0$ &
All $\zeta$ & Stable node & $1$ &-1\\
\hline
\end{tabular}
\end{center}
\caption[crit]{\label{table:crit0}The properties of the critical points for the scalar
for the exponential potential.}
\end{table*}

\begin{figure}
\includegraphics[width=6.5cm]{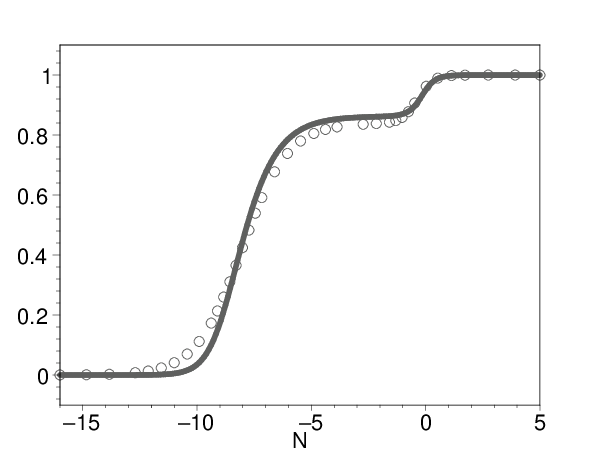}
\caption{ \label{fig:expondensity}The solid line represents the density fraction of the scalar. The circled line represents
the density fraction for the mixture of cold dark matter and cosmological constant in $\Lambda\textrm{CDM}$ model. }
\end{figure}

\begin{figure}
\includegraphics[width=6.5cm]{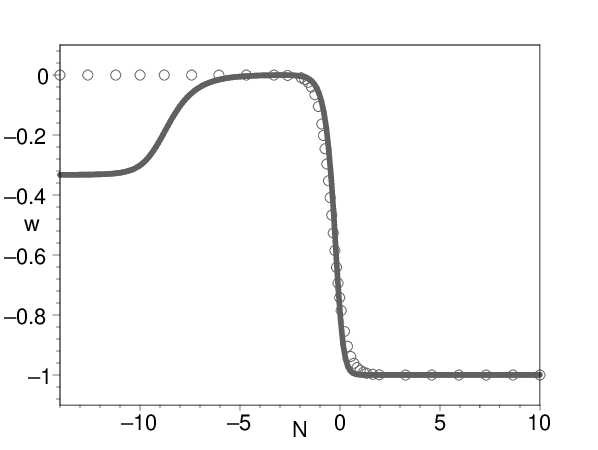}
\caption{\label{fig:exponeos}The solid line represents the evolution of the equation of state for the scalar. The circled line represents
the equation of state for the mixture of cold dark matter and cosmological constant in $\Lambda\textrm{CDM}$ model. }
\end{figure}

In Fig.~\ref{fig:expondensity}, we plot the density fraction of scalar field and the density faction of dark matter and cosmological constant in $\Lambda \textrm{CDM}$ model. The solid line represents the density fraction of the scalar. The circled line represents the density fraction for the mixture of cold dark matter and cosmological constant in $\Lambda\textrm{CDM}$ model. In Fig.~\ref{fig:exponeos}, we plot the equation of state for the scalar and for the mixture of cold dark matter and cosmological constant in $\Lambda \textrm{CDM}$ model. The solid line represents the evolution of the equation of state for the scalar. The circled line represents
the equation of state for the mixture of cold dark matter and cosmological constant in $\Lambda\textrm{CDM}$ model. In both cases, we set $\Omega _{b0}=0.04\;,\ \ \ \Omega_{r0}=8.1\cdot10^{-5}\;,\ \ \ \Omega_{\lambda}=0.75\;$.

\subsection{ Crossing the Phantom Divide}
In the absence of baryon and relativistic matter, we find the scalar can cross the phantom divide ($w=-1$). To demonstrate this, we write down the autonomous
system of equations

\begin{eqnarray}
\label{autoquin109} \frac{dx}{d N}&=&
\frac{\sqrt{6}}{2}\left(1-x^2\right)u+\frac{x}{1+x^2}\left[-2\sqrt{6}x\left(1-x^2\right)u \right.\nonumber
\\
&&\left.-3\left(1-x^2\right)\right]\;,\nonumber\\
\frac{du}{dN}&=&-\sqrt{6}\Gamma xu^2+\sqrt{6}xu^2+\frac{u}{2\left(1+x^2\right)}\left[-3\left(1-x^2\right)\right.\nonumber
\\
&&\left.-2\sqrt{6}x\left(1-x^2\right)u\right]+\frac{3u}{2}\;,
\end{eqnarray}
together with the constraint equation
\begin{equation}
x^2+y^2=1\,.
\end{equation}
The equation of state $w$ is given by
\begin{eqnarray}
w&\equiv&\frac{p_\phi}{\rho_\phi}=\frac{1}{3}\cdot\frac{2\sqrt{6}ux^3-2\sqrt{6}xy-3+3x^2}{1+x^2}\;.
\end{eqnarray}

In Table~\ref{table:critcrossing}, we present the properties of the critical points
for the exponential potential. The points (a) corresponds to the
kinetic energy dominated epoch and this point is unstable. In this epoch, the scalar has the
equation of state $0$. The points (b)
corresponds to the epoch where both the kinetic energy and potential energy vanishes, and this point is stable and an attractor.
In this epoch, the scalar field behaves as the cosmological constant.

\begin{figure}
\includegraphics[width=6.5cm]{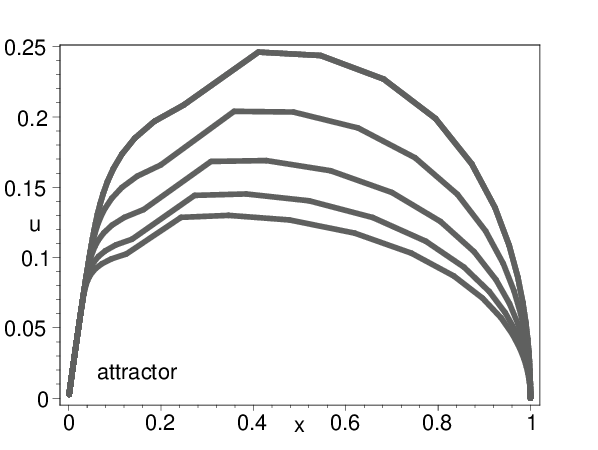}
\caption{\label{fig:exponcrossing}The phase plane for the exponential potential. The point (1, 0) corresponds to the
kinetic energy dominated epoch and it is unstable. The
point (0 ,0) corresponds to the
potential energy dominated epoch and it is stable. These trajectories show that the scalar
always evolves from the kinetic energy dominated epoch to the potential energy dominated epoch. }
\end{figure}

\begin{figure}
\includegraphics[width=6.5cm]{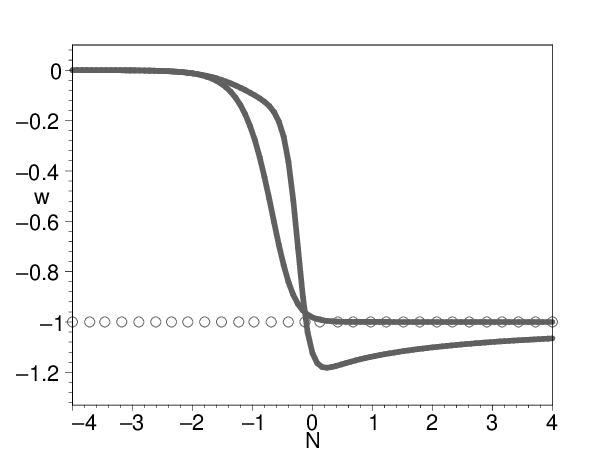}
\caption{\label{fig:eoscrossing}The equation of state for the scalar with two different initial conditions $x(0)=0.1, u(0)=0.1$ and $x(0)=0.1, u(0)=0.9$. For the condition $x(0)=0.1, u(0)=0.9$, the scalar can across the phantom divide.}
\end{figure}

In Fig.~\ref{fig:exponcrossing} and Fig.~\ref{fig:eoscrossing}, we plot the phase plane and the behavior of the equation of state during phantom crossing. Although it can cross the phantom divide, in the next subsection, we show that the corresponding sound speed of the scalar is either imaginary or superluminal. So this phantom crossing behavior may be physically forbidden.
\begin{table*}[t]
\begin{center}
\begin{tabular}{|c|c|c|c|c|c|c|}
Name &  $x$ & $u$ & Existence & Stability & $\Omega_\phi$
 & $w$ \\
\hline \hline (a) & 1 & 0 & All $\zeta$& Unstable node
& Kinetic Energy Dominated & 0 \\
\hline \hline (b) & 0 & 0 & All $\zeta$& Stable node
&   scalar vanishes & -1 \\
\hline
\end{tabular}
\end{center}
\caption[crit]{\label{table:critcrossing}The properties of the critical points for the scalar
for the exponential potential.}
\end{table*}

\subsection{Sound Speed of the Scalar}

The scalar field is constant in Minkowski spacetime due to $G_{\mu\nu}=0$. So it is trivial to consider the perturbation theory of the field in Minkowski spacetime. Here we consider the field perturbation in
the background of FRW Universe. In this paper, we have assumed the scalar is not coupled to the Standard Model particles but only to gravity. Therefore, in order to derive the equation of motion for the scalar field perturbation, we start from the Lagrangian density:

\begin{eqnarray}
\label{eq:pert}
\mathscr{L}=\sqrt{-g}\left[\frac{1}{18}G^{\mu\nu}\partial_{\mu}\phi\partial_{\nu}\phi+V\left(\phi\right)\right]\;.
\end{eqnarray}
It is convenient to work in Newtonian gauge. In the absence of anisotropic stress, and for scalar perturbations only, the perturbed FRW metric can be written in the form
\begin{eqnarray}
ds^2=-\left(1+2\Phi\right)dt^2+a\left(t\right)^2
\left(1-2\Phi\right)\left(dr^2+r^2d\Omega^2\right)\;,
\end{eqnarray}
where $\Phi$ is the gauge invariant Newtonian potential. The potential
characterizes the metric perturbations.

We proceed to linear order by perturbing the field
\begin{eqnarray}
\phi\left(t\right)\rightarrow \phi\left(t\right)+\delta\phi\left(t,\ \textbf{x}\right)\;.
\end{eqnarray}
We find the explicit form of the perturbation equation has been given in Ref.~\cite{hwang:2001}:
\begin{eqnarray}
\frac{1}{a^3Q}\left(a^3Q\dot{U}\right)^{\cdot}-s\frac{\Delta}{a^2}U=0\;,
\end{eqnarray}
where $\Delta$ is the three dimensional Laplace operator, $s$ is the sound speed squared and
\begin{eqnarray}
&&U\equiv\Phi-\frac{H}{\dot{\phi}}\delta\phi\;,\nonumber\\&&
Q\equiv \frac{3\frac{Q_a^2}{2F+Q_b}+Q_e}{\left(H+\frac{Q_a}{2F+Q_b}\right)^2}\;,\nonumber\\&&
s\equiv 1+\frac{Q_d+\frac{Q_aQ_e}{2F+Q_b}+\left(\frac{Q_a}{2F+Q_b}\right)^2Q_f}{3\frac{Q_a^2}{2F+Q_b}+Q_e}\;,\nonumber\\&&
F\equiv\frac{1}{8\pi}\;,\nonumber\\&&
Q_a\equiv-\frac{2}{9} H\dot{\phi}^2\;,\nonumber\\&&
Q_b\equiv-\frac{1}{9}\dot{\phi}^2\;,\nonumber\\&&
Q_c\equiv\frac{1}{3} H^2\dot{\phi}^2\;,\nonumber\\&&
Q_d\equiv \frac{2}{9}\dot{H}\dot{\phi}^2\;,\nonumber\\&&
Q_e\equiv-\frac{4}{9}\dot{\phi}\left(\ddot{\phi}-H\dot{\phi}\right)\;,\nonumber\\&&
Q_f\equiv-\frac{2}{9}\dot{\phi}^2\;.\nonumber\\&&
\end{eqnarray}
Using the variables defined in the subsection \textrm{II-D}, we find the square of the sound speed can be written as
\begin{eqnarray}
s&=&\frac{1}{3\left(x^2+1\right)\left(x^2-3\right)}\left[9w-30wx^2+9wx^4\right.\nonumber\\&&\left.+
4x\left(x^2-1\right)\left(x+3\sqrt{6}u-\sqrt{6}ux^2\right)\right]\;.
\end{eqnarray}
Then using the autonomous system of equations Eq.~(\ref{autoquin109}), we can plot the evolution of the square of the sound speed.

\begin{figure}
\includegraphics[width=6.5cm]{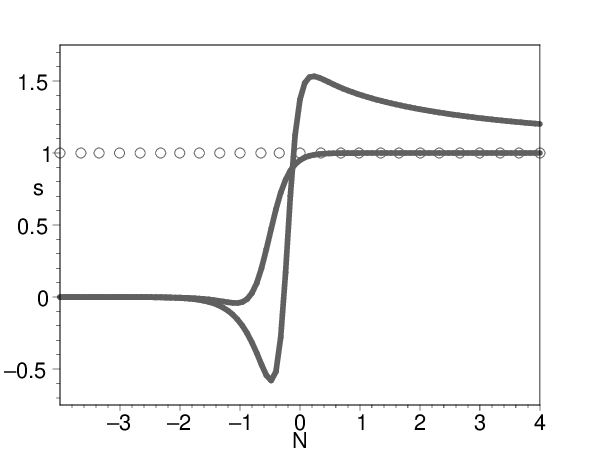}
\caption{\label{fig:speed}Evolution of the sound speed squared $s$ of the scalar with two different initial conditions $x(0)=0.1, u(0)=0.1$ and $x(0)=0.1, u(0)=0.9$. For the condition $x(0)=0.1, u(0)=0.9$, the sound speed $\sqrt{s}$ of the scalar is either imaginary
or superluminal in the vicinity of phantom divide crossing (see Fig.~\ref{fig:eoscrossing}). Therefore, the phantom divide crossing behavior may be physically forbidden.}
\end{figure}

In Fig.~\ref{fig:speed}, we plot the behavior of the sound speed squared. It shows that, in the kinetic term ($G_{\mu\nu}\nabla^{\mu}\phi\nabla^{\nu}\phi$) dominated Universe, the sound speed of scalar always vanishes. Then the vanishing of both pressure and sound speed enable the scalar (without a scalar potential) to be a plausible candidate of cold dark matter. For some initial conditions, for example, $x(0)=0.1,\ u(0)=0.9$, the sound speed would be either imaginary or superluminal in the vicinity of phantom divide crossing (see Fig.~\ref{fig:eoscrossing}). So this behavior may be physically forbidden. But for some initial conditions, for example, $x(0)=0.1,\ u(0)=0.1$, the sound speed of the scalar is always smaller than the speed of light. Therefore, it is viable that the scalar field (with a scalar potential term) is a potential candidate for dark matter and dark energy.

\section{coupled to $n$ Einstein Tensors}
\subsection{Without Potential}
In section II, we investigated the cosmic evolution of the scalar field coupled to a single Einstein tensor. In this section, we shall extend this coupling to $n$ Einstein tensors. Consider the following Lagrangian density of the scalar field, which is coupled to gravitation
 \begin{eqnarray}
\mathscr{L}=\lambda
G^{\mu}_{\alpha_1}G^{\alpha_1}_{\alpha_2}G^{\alpha_2}_{\alpha_3}\cdot\cdot\cdot
G^{\alpha_{n-2}}_{\alpha_{n-1}}G^{\alpha_{n-1}\nu}\partial_{\mu}\phi\partial_{\nu}\phi\;,
\end{eqnarray}
where
\begin{eqnarray}
G^{\mu\alpha_i}=R^{\mu\alpha_i}-\frac{1}{2}g^{\mu\alpha_i}R\;,
\end{eqnarray}
is the Einstein tensor. $\lambda$ is assumed to be a positive
constant. By considering variation of the Lagrangian density with
respect to the field $\phi$, we obtain the following equation of motion
\begin{eqnarray}
\label{eom}
\bigtriangledown_{\mu}\left(G^{\mu}_{\alpha_1}G^{\alpha_1}_{\alpha_2}G^{\alpha_2}_{\alpha_3}\cdot\cdot\cdot
G^{\alpha_{n-2}}_{\alpha_{n-1}}G^{\alpha_{n-1}\nu}\partial_{\nu}\phi\right)=0\;.
\end{eqnarray}
In general, it is a third order differential equation. But in the background of Friedmann-Robertson-Walker (FRW) Universe, it reduces to a second order differential equation.  We also model all other matter sources present in the Universe as perfect fluids. In this case the Einstein equations are given by
\begin{eqnarray}
3H^2&=&8\pi\left(\rho_{\phi}+\rho\right)\;,\nonumber\\
2\dot{H}+3H^2&=&-8\pi\left(p_{\phi}+p\right)\;,
\end{eqnarray}
where the energy density and pressure of the scalar field are given by
\begin{eqnarray}
\rho_{\phi}&=&\lambda\left(2n+1\right)\left(3
H^{2}\right)^{n}\dot{\phi}^2\;,\nonumber\\
p_{\phi}&=&\lambda\left(2n+1\right)\left(3H^2\right)^{n}\dot{\phi}^2\left(1+\frac{2n\dot{H}}{3H^2}\right)\;,
\end{eqnarray}
$\rho,\ p$ are defined by Eq.~(19).

The equation of motion of the scalar field derived from Eq.~(\ref{eom}) takes the form
\begin{eqnarray}
H\ddot{\phi}+\left(2n\dot{H}+3H^2\right)\dot{\phi}=0\;.
\end{eqnarray}
This equation of motion is equivalent to the energy conservation equation of the scalar. From the equation of motion we obtain
\begin{eqnarray}
\dot{\phi}=\frac{c_0}{H^{2n}a^3}\;,
\end{eqnarray}
with $c 0$ an integration constant. This gives us the following Friedmann equation,
\begin{eqnarray}
3H^2=8\pi\left[\rho+\frac{\lambda c_0^2
3^n\left(2n+1\right)}{H^{2n}a^6}\right]\;.
\end{eqnarray}
So the energy density of the scalar field scales as $H^{-2n}a^{-6}$. In the radiation dominated epoch, we have $H^2\sim a^{-4}$, and so the energy density of scalar field would scale as $a^{4n-6}$. It behaves as a phantom field, growing in density with the expansion, when $n\geq 2$.  In the matter dominated epoch $H^2\sim a^{-3}$, so the energy density of scalar field scales as $a^{3n-6}$. It behaves as cosmological constant for $n=2$ and as a phantom field for $n\geq 3$. In total, when $n\geq3$ the scalar behaves as phantom field in over the whole history of the Universe.

Taking into account pressureless matter (dark matter plus baryonic matter) and relativistic matter, we obtain the Friedmann equation in the following dimensionless form
\begin{eqnarray}
h^2=\Omega_{r0}a^{-4}+\Omega_{m0}a^{-3}+\Omega_{X0}a^{-6}h^{-2n}\;,
\end{eqnarray}

\begin{figure}
\includegraphics[width=6.5cm]{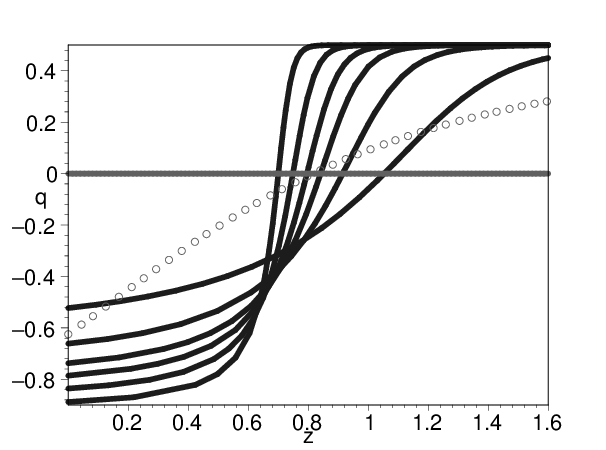}
\caption{\label{fig:qparam}The deceleration parameter for the scalar (solid line)
model and the $\Lambda \textrm{CDM}$ (circled line) model. For the
scalar model, from left to right, we set $n=6,\ 9,\ 12,\ 15,\ 20,\
30$, respectively. }
\end{figure}

In Fig.~\ref{fig:qparam}, we plot the deceleration parameter $q$:
\begin{eqnarray}
q=-\left(1+\frac{\dot{H}}{H^2}\right)\;,
\end{eqnarray}
as a function redshift for different values of $n$. We find the scalar model can really enhance the acceleration of the Universe.

\begin{figure}
\includegraphics[width=6.5cm]{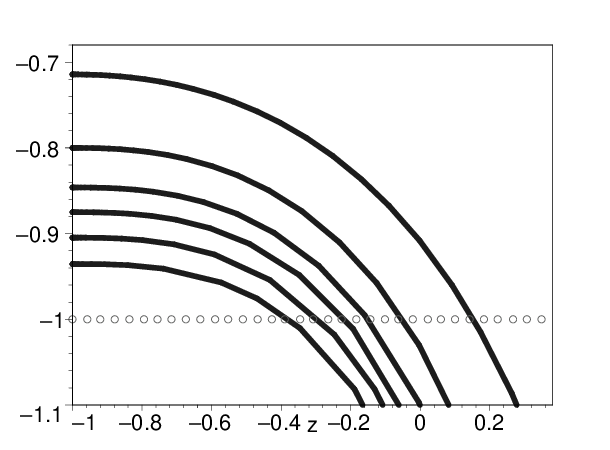}
\caption{\label{fig:ncouplingseos1}The equation of the state of the scalar field as a function of
redshift. From top down, we set $n=6,\ 9,\ 12,\ 15,\ 20,\ 30$,
respectively. They smoothly cross the phantom divide.}
\end{figure}

\begin{figure}
\includegraphics[width=6.5cm]{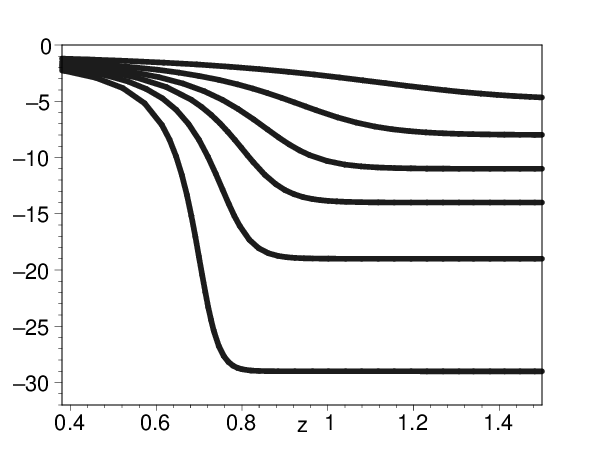}
\caption{\label{fig:ncouplingseos2}The equation of the state of the scalar field as a function of
redshift. From top down, we set $n=6,\ 9,\ 12,\ 15,\ 20,\ 30$,
respectively. They asymptotically approach the constant equation of
state at higher redshifts. }
\end{figure}

\begin{figure}
\includegraphics[width=6.5cm]{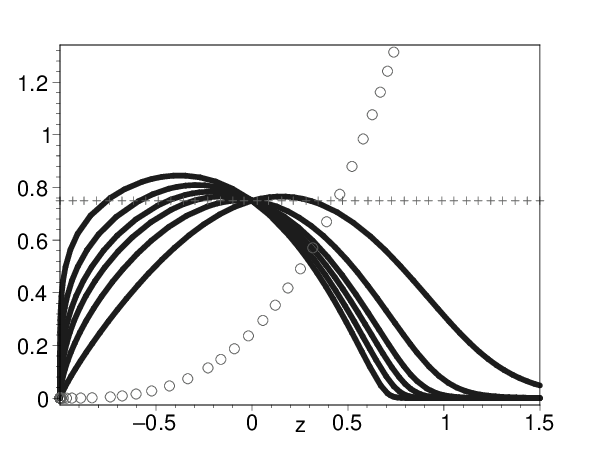}
\caption{\label{fig:ncouplingsdens}The energy density $\Omega_{X0}a^{-6}h^{-2n}$ of the scalar
field (solid line) and the energy density $\Omega {m0}a^{-3}$ of
matter (circled line). The crossed line is for the cosmological
constant. From right to left, we set $n=6,\ 9,\ 12,\ 15,\ 20,\ 30$,
respectively. They all asymptotically approach zero at higher
redshifts. For $n\geq 9$, the density of the scalar is vanishing at
redshifts greater than unit one. }
\end{figure}

In Fig.~\ref{fig:ncouplingseos1} and Fig.~\ref{fig:ncouplingseos2}, we plot the equation of state for the scalar.
In Fig.~\ref{fig:ncouplingsdens}, we plot the energy density $\Omega_{X0}a^{-6}H^{-2n}$ of
the scalar and the energy density $\Omega_{m0}a^{-3}$ of matter,
respectively.  For different values of $n$, the energy density is always asymptotically approaching zero at higher
redshifts. For $n\geq 9$, the density is vanishing at
redshifts greater than one. Thus by simply choosing a relative large $n$ we obtain a dark energy negligible at redshifts greater than one without fine tuning of initial conditions. We know the energy density of the usual phantom field grows during most of the history of the Universe, so adjusting it to become important very recently requires even more fine tuning of the initial conditions. However, in above models, for a very large range values  of $n$ ($n>9$), the scalar field becomes important very recently, as shown in Fig.~\ref{fig:ncouplingseos2}.

\subsection{With Potential}
In this section, we investigate the cosmic behavior of the scalar field in the presence of a scalar potential.
The Lagrangian density is given by
 \begin{eqnarray}
\mathscr{L}&=&\frac{1}{2\left(2n+1\right)3^{n}}
G^{\mu}_{\alpha_1}G^{\alpha_1}_{\alpha_2}\cdot\cdot\cdot
G^{\alpha_{n-1}\nu}\partial_{\mu}\phi\partial_{\nu}\phi\nonumber\\&&+V\left(\phi\right)\;,
\end{eqnarray}
where $n$ a positive integer. Then the Einstein equations are given by
\begin{eqnarray}
3H^2&=&8\pi\left(\rho_{\phi}+\rho\right)\;,\nonumber\\
2\dot{H}+3H^2&=&-8\pi\left(p_{\phi}+p\right)\;,
\end{eqnarray}
where the energy density and pressure of the scalar field are given
by
\begin{eqnarray}
\rho_{\phi}&=&
\frac{1}{2}H^{2n}\dot{\phi}^2+V\;,\nonumber\\
p_{\phi}&=&\frac{1}{2}H^{2n}\dot{\phi}^2\left(1+\frac{2n\dot{H}}{3H^2}\right)+\frac{2n}{3}V^{'}\dot{\phi}H^{-1}-V\;.
\end{eqnarray}
The equation of motion of the scalar field derived from Eq.~(\ref{eom}) takes the form
\begin{eqnarray}
\ddot{\phi}+3H\dot{\phi}\left(1+\frac{2n\dot{H}}{3H^2}\right)+\frac{2n+1}{H^{2n}}V^{'}=0\;.
\end{eqnarray}
We introduce the following dimensionless quantities
\begin{eqnarray}
&& x\equiv\sqrt{\frac{4\pi}{3}}H^{n-1}\dot{\phi}\;,\ \ \ y\equiv\sqrt{\frac{8\pi V}{3H^2}}\;,\ \ \ z\equiv\sqrt{\frac{8\pi \rho_m}{3H^2}}\;,\ \ \ \nonumber\\&&u\equiv-\frac{V^{'}}{\sqrt{8\pi}VH^{n}}\;,\ \ \ \Gamma\equiv\frac{V^{''}V}{V^{'2}}\;,\ \ \   N\equiv\ln a\;.
\end{eqnarray}
Then the above equations can be written in the following autonomous form
\begin{eqnarray}
\label{autoquin1} \frac{dx}{d N}&=&
\frac{\sqrt{6}}{2}\left(2n+1\right)y^2u-3x+\frac{\left(n+1\right)x}{2\left(1+nx^2\right)}\left[4+2x^2\right.\nonumber
\\
&&\left.-4y^2-z^2-2\sqrt{6}nxy^2u\right]\;,\nonumber\\
\frac{dy}{d N} &=&
-\frac{\sqrt{6}}{2}xyu+\frac{y}{2\left(1+nx^2\right)}\left[4+2x^2\right.\nonumber
\\
&&\left.-4y^2-z^2-2\sqrt{6}nxy^2u\right]\,,\nonumber\\
\frac{dz}{dN}&=&-\frac{3}{2}z+\frac{z}{2\left(1+nx^2\right)}\left[4+2x^2\right.\nonumber
\\
&&\left.-4y^2-z^2-2\sqrt{6}nxy^2u\right]\;,\nonumber\\
\frac{du}{dN}&=&-\sqrt{6}\Gamma xu^2+\sqrt{6}xu^2+\frac{nu}{2\left(1+nx^2\right)}\left[4+2x^2\right.\nonumber
\\
&&\left.-4y^2-z^2-2\sqrt{6}nxy^2u\right]\;,
\end{eqnarray}
together with a constraint equation
\begin{equation}
x^2+y^2+z^2+\frac{8\pi \rho_r}{3H^2}=1\,.
\end{equation}
The equation of state $w$ and the
fraction of the energy density $\Omega_{\phi}$ for the scalar field
are
\begin{eqnarray}
w&\equiv&\frac{p_\phi}{\rho_\phi}=\frac{{x^2}}{\left(1+nx^2\right)\left(3x^2+3y^2\right)}\left[2\sqrt{6}n^2xy^2u
+3\right.\nonumber
\\
&&\left.+nx^2-4n+4ny^2+nz^2\right]-\frac{2\sqrt{6}nxy^2u+3y^2}{3x^2+3y^2}\;,\nonumber\\
\Omega_{\phi}&\equiv&\frac{8\pi\rho_\phi}{3H^2}=x^2+y^2\;,\nonumber\\
q&\equiv&-\left(1+\frac{\dot{H}}{H^2}\right)\nonumber\\&&=-1+\frac{4+2x^2-4y^2-z^2-2\sqrt{6}nuxy^2}{2\left(1+nx^2\right)}
\end{eqnarray}
As an example, we have considered the exponential potential $V=e^{-\zeta\phi}$ with $\zeta$ a positive constant.

In Table \ref{tab:ncouplingscrit}, we present the properties of the critical points
for the exponential potential. The point (a) corresponds to the
relativistic matter dominated epoch and this point is unstable. In this epoch, the scalar has the
equation of state $w=1-\frac{4n}{3}$. The point (b)
corresponds to the kinetic energy of the scalar dominated epoch and is a saddle point.
In this epoch, the scalar field has the equation of state $w=\frac{1-n}{1+n}$. The point (c) corresponds to the potential energy dominated epoch. It is stable and an attractor. In this epoch, the scalar has the equation of state $w=-1$. The point (d) corresponds to the matter
dominated epoch. It is a saddle point. In the
epoch of (d), the scalar has the equation of state $1-n$.

We note that when $n=1$, the scalar behaves as:
a curvature term in radiation epoch; pressureless matter in both the kinetic energy dominated epoch and the matter dominated epoch;
a cosmological constant in potential energy dominated epoch. When $n=2$, the equation of state is smaller than $-1/3$.
\begin{table*}[t]
\begin{center}
\begin{tabular}{|c|c|c|c|c|c|c|c|c|}
Name &  $x$ & $y$ & $z$ & $u$ & Existence & Stability & $\Omega_\phi$
 & $w$ \\
\hline \hline (a) & 0 & 0 & 0 & 0 & All $\zeta$& Unstable node
&   0 & $1-\frac{4n}{3}$\\
\hline (b) & $1$ & 0 & 0 & 0 & All $\zeta$ & Saddle line & $1$ & $\frac{1-n}{1+n}$  \\
\hline (c) & 0&1
& 0& $0$ &
All $\zeta$ & Stable node & $1$ &$-1$\\
\hline (d) & 0&0
& 1& $0$ &
All $\zeta$ & Saddle node & $0$ &$1-n$\\
\hline
\end{tabular}
\end{center}
\caption[crit]{\label{tab:ncouplingscrit}The properties of the critical points for the scalar
for the exponential potential.}
\end{table*}

\begin{figure}
\includegraphics[width=6.5cm]{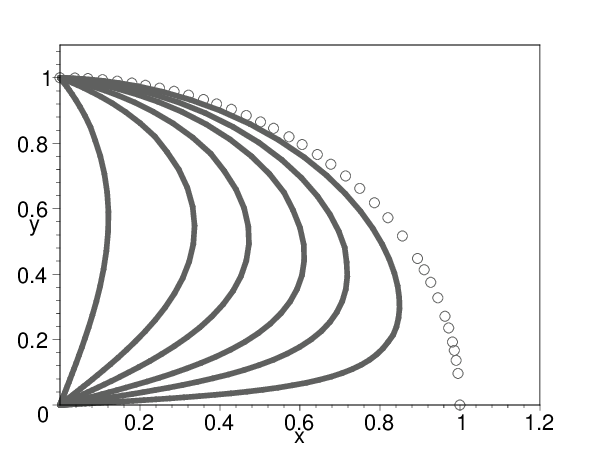}
\caption{\label{fig:ncouplingsphase}The phase plane for the exponential potential with $n=2$. The point (0, 0) corresponds to the
radiation dominated epoch and the circled arc ($x^2+y^2=1$) corresponds to
the scalar dominated epoch. The point (0, 0) is unstable and the
point (0 ,1) is stable and thus an attractor. These trajectories show that the Universe
always evolves from the radiation dominated epoch to the scalar
field dominated epoch and ends at the scalar potential dominated epoch.}
\end{figure}

In Fig.~\ref{fig:ncouplingsphase}, we plot the phase portraits for the scalar with vast
initial conditions. The point (0, 0) corresponds to the
radiation dominated epoch and the circled arc ($x^2+y^2=1$) corresponds to
the scalar dominated epoch. The point (0, 0) is unstable and the
point (0 ,1) is stable and an attractor. These trajectories show that the Universe
always evolves from the radiation dominated epoch to the scalar
field dominated epoch and ends at the scalar potential dominated epoch.

\begin{figure}
\includegraphics[width=6.5cm]{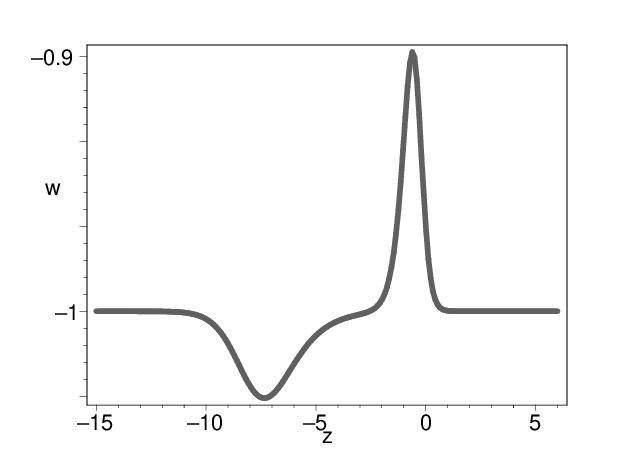}
\caption{\label{fig:ncouplingsevoleos}Evolution of the equation of state for the scalar. The scalar can cross the phantom divide. But the equation of state $w\simeq-1$ over most of the history of the Universe.}
\end{figure}

\begin{figure}
\includegraphics[width=6.5cm]{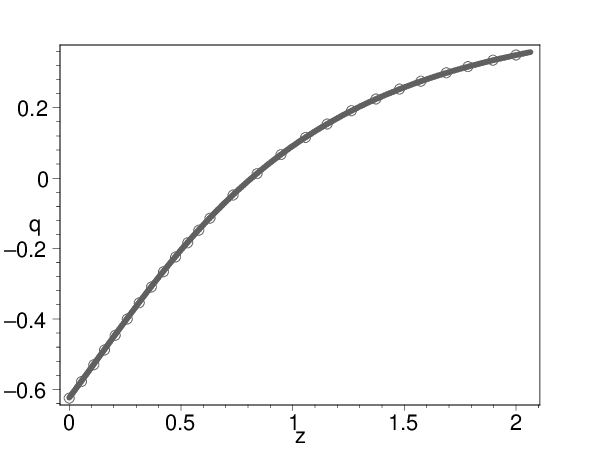}
\caption{\label{fig:ncouplingsdecelparam}The evolution of decelerating parameters for $\Lambda
\textrm{CDM}$ model (pointed line) and the scalar model (solid line).
Both models predict the transition redshift of the Universe from
deceleration to acceleration at $z T\simeq0.8$.}
\end{figure}

In Fig.~\ref{fig:ncouplingsevoleos}, we plot the equation of state for the scalar when $n=2$. As shown in Table \ref{tab:ncouplingscrit},
the equation of state is smaller than $-1/3$. Buy choosing some smaller $\zeta$, we have the equation of state $w\simeq-1$
in the total history of the Universe. In Fig.~\ref{fig:ncouplingsdecelparam}, we plot the deceleration of the Universe for the scalar model and $\Lambda \textrm{CDM}$ model.
We find the two models predict nearly the same behavior of the Universe from deceleration to acceleration.
This is because the equation of state for dark energy is $w\simeq -1$ over the entire history of the Universe, and so the energy density of this dark energy is nearly a constant.

\subsection{New Inflaton field}
In this section, we investigate the cosmic behavior of the scalar field in the absence of other matter sources.
In this case, we have the following autonomous system of equations
\begin{eqnarray}
\label{autoquin1} \frac{dx}{d N}&=&
\frac{\sqrt{6}}{2}\left(2n+1\right)\left(1-x^2\right)u-3x+\frac{\left(n+1\right)x}{2\left(1+nx^2\right)}\left[4+2x^2\right.\nonumber
\\
&&\left.-4\left(1-x^2\right)-2\sqrt{6}nx\left(1-x^2\right)u\right]\;,\nonumber\\
\frac{du}{dN}&=&-\sqrt{6}\Gamma xu^2+\sqrt{6}xu^2+\frac{nu}{2\left(1+nx^2\right)}\left[4+2x^2\right.\nonumber
\\
&&\left.-4\left(1-x^2\right)-2\sqrt{6}nx\left(1-x^2\right)u\right]\;,
\end{eqnarray}
together with a constraint equation
\begin{equation}
x^2+\frac{8\pi V}{3H^2}=1\,.
\end{equation}
The equation of state $w$ is
\begin{eqnarray}
w&\equiv&\frac{p_\phi}{\rho_\phi}=\frac{{x^2}}{3\left(1+nx^2\right)}\left[2\sqrt{6}n^2x\left(1-x^2\right)u
+3\right.\nonumber
\\
&&\left.-3nx^2\right]-\frac{2\sqrt{6}}{3}nx\left(1-x^2\right)u-1\;,
\end{eqnarray}
As an example, we consider the exponential potential $V=e^{-\zeta\phi}$ with $\zeta$ a positive constant.

In Table \ref{fig:ncouplingsscalarpotcrit}, we present the properties of the critical points
for the exponential potential. The points (a) corresponds to the
kinetic energy dominated epoch and this point is unstable. In this epoch, the scalar has the
equation of state $\frac{1-n}{1+n}$. The points (b)
corresponds to the scalar vanishes and this point is stable and an attractor.
In this epoch, the scalar field also has the equation of state $-1$. So by choosing a large $n$ we find
the equation of state is always approximately equal to $-1$, without fine tuning the initial conditions.
For the inflaton field, one generally assumes that the scalar potential is sufficiently flat to make the equation of state
$w\simeq -1$. In this case, our scalar field can easily achieve this equation of state without the flatness condition. Therefore this scalar field could act as a new candidate for the inflaton.
\begin{table*}[t]
\begin{center}
\begin{tabular}{|c|c|c|c|c|c|c|}
Name &  $x$ & $u$ & Existence & Stability & $\Omega_\phi$
 & $w$ \\
\hline \hline (a) & 1 & 0 & All $\zeta$& Unstable node
& Kinetic Energy Dominated & $\frac{1-n}{1+n}$ \\
\hline \hline (b) & 0 & 0 & All $\zeta$& Stable node
&  scalar vanishes  & -1 \\
\hline
\end{tabular}
\end{center}
\caption[crit]{\label{fig:ncouplingsscalarpotcrit}The properties of the critical points for the scalar
with an exponential potential} 
\end{table*}

\begin{figure}
\includegraphics[width=6.5cm]{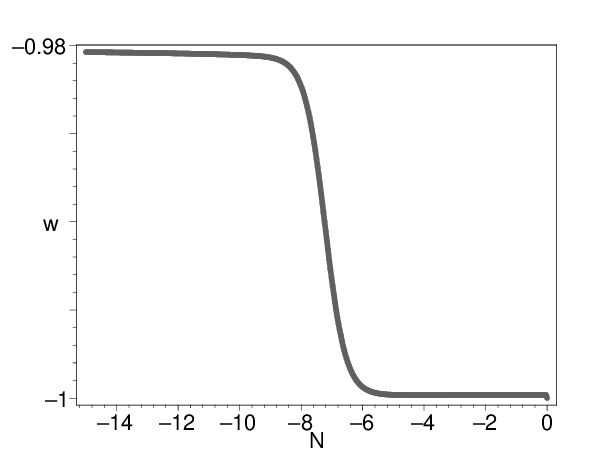}
\caption{\label{fig:100couplingseos}The equation of state for $n=100$.}
\end{figure}

In Fig.~\ref{fig:100couplingseos}, we plot the equation of state for $n=100$.

\section{conclusion and discussion}
In this paper, we have investigated the cosmic evolution of the
scalar field with the kinetic term coupled to $n$ Einstein tensors.
When $n=1$, the equation of motion is a second order differential equation, and so is a good theory. We find that the scalar has some interesting properties. Firstly, in the absence of other matter sources
or in the presence of only pressureless matter, the scalar (without a scalar potential) behaves
exactly as pressureless matter. Further more, the sound speed of the scalar perturbations is exactly zero in the structure formation era.
These properties enable the scalar to be a potential candidate for cold
dark matter. Secondly, in the presence of a scalar potential, the scalar field has the equation of state $-1\leq w\leq 0$.
So the scalar may play the role of both cold dark matter and dark
energy. It is found the sound speed is always smaller than the speed of light, and so it is physically viable.
By investigating the dynamics of the scalar field in the absence of other matter sources,
we find it can cross the phantom divide. But it may be physically forbidden due to the presence of a superluminal sound speed.
Finally, if the kinetic term is coupled to more than one Einstein tensors, the equation of state is always approximately equals to $-1$ whether the potential is flat or not. Thus the scalar may also be a potential candidate for the inflaton field.

\acknowledgments

We would like to thank the anonymous referee for the insightful
comments and suggestions, which have allowed us to improve this paper significantly. We especially  thank Andrew R. Liddle, Martin Kunz and David Parkinson for stimulating and
illuminating discussions. We are grateful to David Parkinson for the English improvement of this paper. This work is
supported by the National Science Foundation of China under the
Key Project Grant 10533010, Grant 10575004, Grant
10973014, and the 973 Project (No. 2010CB833004).

As we were completing this work, L. N. Granda \cite{granda:2009}, E. N. Saridakis and S. V. Sushkov \cite{sushkov:2010} arXived two papers considering part of the points in this work.

\newcommand\AL[3]{~Astron. Lett.{\bf ~#1}, #2~ (#3)}
\newcommand\AP[3]{~Astropart. Phys.{\bf ~#1}, #2~ (#3)}
\newcommand\AJ[3]{~Astron. J.{\bf ~#1}, #2~(#3)}
\newcommand\APJ[3]{~Astrophys. J.{\bf ~#1}, #2~ (#3)}
\newcommand\APJL[3]{~Astrophys. J. Lett. {\bf ~#1}, L#2~(#3)}
\newcommand\APJS[3]{~Astrophys. J. Suppl. Ser.{\bf ~#1}, #2~(#3)}
\newcommand\JHEP[3]{~JHEP.{\bf ~#1}, #2~(#3)}
\newcommand\JCAP[3]{~JCAP. {\bf ~#1}, #2~ (#3)}
\newcommand\LRR[3]{~Living Rev. Relativity. {\bf ~#1}, #2~ (#3)}
\newcommand\MNRAS[3]{~Mon. Not. R. Astron. Soc.{\bf ~#1}, #2~(#3)}
\newcommand\MNRASL[3]{~Mon. Not. R. Astron. Soc.{\bf ~#1}, L#2~(#3)}
\newcommand\NPB[3]{~Nucl. Phys. B{\bf ~#1}, #2~(#3)}
\newcommand\PLB[3]{~Phys. Lett. B{\bf ~#1}, #2~(#3)}
\newcommand\PRL[3]{~Phys. Rev. Lett.{\bf ~#1}, #2~(#3)}
\newcommand\PR[3]{~Phys. Rep.{\bf ~#1}, #2~(#3)}
\newcommand\PRD[3]{~Phys. Rev. D{\bf ~#1}, #2~(#3)}
\newcommand\RMP[3]{~Rev. Mod. Phys.{\bf ~#1}, #2~(#3)}
\newcommand\SJNP[3]{~Sov. J. Nucl. Phys.{\bf ~#1}, #2~(#3)}
\newcommand\ZPC[3]{~Z. Phys. C{\bf ~#1}, #2~(#3)}
\newcommand\IJGMMP[3]{~Int. J. Geom. Meth. Mod. Phys{\bf ~#1}, #2~(#3)}
\newcommand\CTP[3]{~Commun. Theor. Phys.{\bf ~#1}, #2~(#3)}
\newcommand\CQG[3]{~ Class. Quant. Grav.{\bf ~#1}, #2~(#3)}
\newcommand\MPLA[3]{~ Mod. Phys. Lett. A{\bf ~#1}, #2~(#3)}

\end{document}